\documentclass[11pt]{JHEP3}
\usepackage{amsmath,amssymb,epsfig}
\usepackage{amsfonts}
\usepackage{verbatim}
\usepackage{latexsym}
\usepackage{amsthm}
\usepackage{amsbsy}

\def\one{{\,\hbox{1\kern-.8mm l}}}
\newcommand{\Dslash}{\not{\hbox{\kern-4pt $D$}}}
\newcommand{\pdslash}{\not{\hbox{\kern-2pt $\partial$}}}

\newcommand{\sign}{\mathrm{sign}}

\newcommand{\La}{\Lambda}

\newcommand{\Comment}[1]{{}}
\newcommand{\diag}{{\rm diag}}

\newcommand{\Aone}{A^{(1)}}
\newcommand{\Atwo}{A^{(2)}}
\newcommand{\stardual}{{{}^*\!}}
\newcommand{\eff}{{\rm eff}}

\def\IZ{{\mathbb Z}}

\def\IR{{\mathbb R}}

\newcommand{\gYM}{{g_{\!\it YM}}}

\newcommand{\bc}{\begin{center}}
\newcommand{\ec}{\end{center}}
\newcommand{\ba}{\begin{array}}
\newcommand{\ea}{\end{array}}
\newcommand{\beq}{\begin{equation}}
\newcommand{\eeq}{\end{equation}}
\newcommand{\bea}{\begin{eqnarray}}
\newcommand{\eea}{\end{eqnarray}}
\newcommand{\bmx}{\begin{pmatrix}}
\newcommand{\emx}{\end{pmatrix}}
\newcommand{\nn}{\nonumber}

\newcommand{\be}{\begin{equation}}
\newcommand{\ee}{\end{equation}}

\newcommand{\del}{\partial}
\newcommand{\half}{{\frac{1}{2}\,}}
\newcommand{\tr}{{\rm tr}}

\newcommand{\eref}[1]{Eq.\,(\ref{#1})}

\newcommand{\vq}{{\vec{q}}}

\newcommand{\vV}{{\vec{V}}}
\newcommand{\vW}{{\vec{W}}}

\newcommand{\vv}{{\vec{v}}}

\newcommand{\tA}{{\tilde A}}

\newcommand{\tF}{{\tilde F}}

\newcommand{\tLambda}{{\tilde \Lambda}}
\newcommand{\cG}{{\cal G}}

\newcommand{\DC}{D^{(C)}}
\newcommand{\FC}{F^{(C)}}

\def\IB{\relax{\rm I\kern-.18em B}}
\def\IC{{\relax\hbox{\kern.3em{\cmss I}$\kern-.4em{\rm C}$}}}
\def\ID{\relax{\rm I\kern-.18em D}}
\def\IE{\relax{\rm I\kern-.18em E}}
\def\IF{\relax{\rm I\kern-.18em F}}
\def\II{\relax{\rm I\kern-.18em I}}
\def\IZ{\relax{\sf Z\kern-.35em Z}}
\def\Id{\relax{1\kern-.32em 1}}
\def\IG{\relax\hbox{$\inbar\kern-.3em{\rm G}$}}
\def\IR{\relax{\rm I\kern-.18em R}}
\newcommand\sfrac[2]{{\textstyle\frac{#1}{#2}}}

\newcommand\shalf{{\textstyle\frac12}}

\normalsize

\title{Unravelling the novel Higgs mechanism in (2+1)d 
Chern-Simons theories}

\author{Sunil Mukhi\footnote{E-mail: mukhi@tifr.res.in}\\~\\
{\it Tata Institute of Fundamental Research,\\ 
Homi Bhabha Rd, Mumbai 400 005, India}}

\abstract{Chern-Simons gauge theories in 2+1 dimensions with multiple
  gauge fields exhibit novel properties that are analysed here in some
  detail. A striking feature is the possibility of a non-propagating
  Chern-Simons field acquiring a {\em massless} propagating mode via a Higgs
  mechanism. This novel Higgs mechanism, originally discovered in the
  context of M-theory, is studied here without reference to
M-theory or supersymmetry. It is revealed as a variant of
  topological mass generation and shown to arise only
  when Chern-Simons and mass matrices are not simultaneously
  diagonalisable. Sufficient conditions are found for it to occur.  It
  is speculated that some analogue of the NHM could occur in 
theories of condensed-matter systems similar to 
those exhibiting the fractional quantum Hall effect, as well as in
2+1 dimensional gravity.} 

\preprint{TIFR/TH/11-45}


\begin{document}

\section{Introduction}

Three-dimensional gauge theories have some special features, notably
the possibility of writing a Chern-Simons kinetic term that is
first-order in the 
derivatives\cite{Jackiw:1980kv,Schonfeld:1980kb,Deser:1981wh}. 
In addition to this
term, a (2+1)d gauge theory can have a conventional Maxwell kinetic
term as well as an explicit mass term that typically arises via a
Higgs mechanism. The possibility of three distinct types of terms in
the Lagrangian for free gauge fields makes for a more subtle spectrum
than is familiar in four-dimensional field theory, where a free gauge
field can only have a Maxwell or mass term.  In particular, parity
violation is a generic feature of theories with a Chern-Simons term.

If one now extends this structure to multiple species of gauge fields,
new subtleties arise. One of the most striking of these is
a new type of Higgs mechanism wherein a non-propagating gauge field
absorbs the degree of freedom of a Higgs field and turns into a {\em
  massless} propagating gauge field\cite{Mukhi:2008ux}. This
phenomenon, the ``novel Higgs mechanism'' (NHM), was discovered and
usefully applied to study the moduli space of extended superconformal
field theories describing multiple membranes in M-theory, namely the
BLG and ABJM field theories. Despite its origin, the mechanism itself
does not rely on supersymmetry or string/M-theory. It is a subtle but
elementary feature of ordinary quantum field theory in (2+1)d, and as
such could be of interest in a much wider context. In particular, it
might well have applications in condensed-matter systems, where
Chern-Simons field theories are already known to play an interesting
role (see for example 
\cite{Semenoff:1988jr,Zhang:1988wy,Read:1988mp,Wen:1991xx}).
Notably the NHM occurs in systems that conserve
parity\cite{Bandres:2008vf,VanRaamsdonk:2008ft}. 

One purpose of this note is to analyse the spectrum of (2+1)d gauge
theories with multiple gauge fields. We will see that the novel Higgs
mechanism arises for a very general class of such theories. Our
analysis reveals that the NHM arises out of a conflict between
simultaneous diagonalisability of the kinetic (Chern-Simons) terms and
the mass terms. This conflict in turn arises when one has Chern-Simons
terms of both signs. There is no analogue of this phenomenon in
``normal'' field theories with second-order kinetic terms, whose sign
is determined by requiring the absence of
negative-norm states.  One consequence of non-diagonalisability in
(2+1)d is that it is not straightforward to read off the spectrum. As
a result the spectrum can have various unusual features and the NHM
turns out to be just one of these.

A well-known characteristic of (2+1)d gauge theories with a single
gauge field, observed nearly three decades ago\cite{Deser:1981wh}, is
that Maxwell gauge fields in 2+1 dimensions acquire a ``topological
mass'' when a Chern-Simons interaction is added to the action. The
propagating degrees of freedom have a single degree of freedom with
spin $+1$ but no corresponding spin $-1$ state (the reverse holds if
we change the sign of the mass parameter).
Subsequently\cite{Deser:1984kw}, it was noticed that the topologically
massive action, containing Maxwell and Chern-Simons terms, is
equivalent to a different action with a Chern-Simons term and an
explicit mass term\cite{Townsend:1983xs} but no Maxwell term. Both
sides of this equivalence correspond to massive theories. We will see
that the analogue of this duality in systems with multiple gauge
fields provides a natural setting to understand the NHM. Namely, in
appropriately chosen systems the analogous duality maps a set of
Chern-Simons and mass terms to a Maxwell term {\em without} a mass.
In turn, the original theory with Chern-Simons and mass terms in turn
can descend from a conformal theory in which the mass terms are
induced by a Higgs mechanism. This means a massless propagating
Yang-Mills field can arise on the Higgs branch of a conformal field
theory.

In the context of M-theory membranes the NHM was originally
presented\cite{Mukhi:2008ux} in a non-Abelian form and the emphasis
was on those features that are relevant to multiple membrane dynamics.
However in the present work we first spell it out in an Abelian
setting with Higgs fields carrying U(1) charges. Subsequently we
introduce non-Abelian gauge fields with corresponding Yang-Mills and
non-Abelian Chern-Simons terms, and discuss the role of interactions.
Towards the end some possible applications of the results are briefly
explored.

\section{Topological mass and novel Higgs mechanism}

\subsection{Topologically massive theories}

We start by reviewing the concept of topologically massive gauge
theories in 2+1 dimensions\cite{Deser:1981wh} and their
duality\cite{Deser:1984kw} to a class of Chern-Simons theories with an
explicit mass term, proposed in Ref.\cite{Townsend:1983xs} where they
were dubbed ``self-dual actions''. It will be seen that the novel
Higgs mechanism is an extension of these theories and dualities to the
case of multiple gauge fields, providing a natural setting for the
phenomenon. Since the immediate goal is to understand the spectrum of
excitations, we initially work at the linearised level, or
equivalently in the Abelian theory.

The basic topologically massive theory is given by a Lagrangian for a
1-form Abelian gauge field $A$ with the Maxwell + Chern-Simons Lagrangian:
\be
{\cal L}_1=\shalf dA\wedge\stardual dA -\shalf m A\wedge dA
\label{atmac}
\ee
The equations of motion of this theory are:
\be
d\,\stardual dA = m\, dA
\label{atmeq}
\ee
As explained in Ref.\cite{Deser:1981wh}, this theory has a single
on-shell degree of freedom that is massive and has spin $+1$.  Because
parity is violated, it is possible to have no spin $-1$ mode. 

Next consider a different Lagrangian consisting of a Chern-Simons term plus an
explicit mass term:
\be
{\cal L}_2=\shalf A\wedge dA + \shalf m\, A\wedge\stardual A
\label{asdac}
\ee
This is the self-dual Lagrangian of Ref.\cite{Townsend:1983xs}. 
This time the equations of motion are:
\be
\stardual dA = mA
\label{asdeq}
\ee
It is easy to verify that the theory of \eref{asdac} is equivalent to
that of \eref{atmac}.  Classically, the equivalence is shown as
follows. First, 
\be 
\stardual dA = mA~\implies~ d\,\stardual dA = m\,dA 
\ee 
so ${\cal L}_2\implies {\cal L}_1$.  For the converse, 
\be
d\,\stardual dA = m\, dA \implies d\left(\stardual dA - m\, A\right)=0
\implies \stardual dA - m\, A=d\lambda
\label{converse}
\ee
and a field re-definition $A\to A-\sfrac{1}{m}d\lambda$ gives
\eref{asdeq}. Thus ${\cal L}_1\implies {\cal L}_2$.  A detailed
discussion of the quantum-mechanical equivalence of the two theories
can be found in Ref.\cite{Deser:1984kw}.

Let us make some comments on  this equivalence.

1. Comparing the two Lagrangians \eref{atmac} and \eref{asdac} we see
that the former is gauge-invariant while the latter does not seem to
have a gauge symmetry. In fact, $A$ in the second Lagrangian can be
thought of as the dual field strength $\stardual dA$ of the first one,
so the second Lagrangian actually expresses the field content of the
first one in gauge-invariant variables. This is seen more clearly 
by introducing\cite{Deser:1984kw} a
``master Lagrangian'' involving two independent 1-form fields $A$ and
$f$: 
\be 
{\cal L}_{\rm master} = \shalf f\wedge \stardual f +f\wedge dA
- \shalf m A\wedge dA
\label{amaster}
\ee 
The equation of motion of $f$ gives:
\be
f=\stardual dA
\ee
while the equation of motion of $A$ is:
\be
df=m\,dA
\ee
Eliminating $f$ we recover \eref{atmeq}, while eliminating $A$ (it is
not algebraic, but only appears as $dA$ so it can still be eliminated
between the two equations of motion) we find \eref{asdeq} for the
field $f$. 

2. While \eref{atmac}
has a smooth massless limit, \eref{asdac} becomes purely topological
and thereby loses a degree of freedom as $m\to 0$. Indeed the proof of
equivalence between the two theories involves a gauge transformation
that becomes singular as $m\to 0$ (see below \eref{converse}). 

3. In (2+1)d, the sign of the mass $m$ of a state of a spin-1 field is
meaningful and can be interpreted as follows: $m>1$ means the state
has spin $+1$ and $m<1$ means it has spin $-1$. For $m=0$ the little
group is trivial and there is no concept of spin.

4. As observed in Ref.\cite{Deser:1989gf}, in addition to gauge fields
one can imagine having a scalar Higgs field in the system such that
its vacuum expectation value generates the mass term in \eref{asdac}.
Therefore the conversion of a Chern-Simons action into a topologically
massive Yang-Mills action can be viewed as a kind of Higgsing.

5. The coefficient of an interacting Chern-Simons term corresponding
to a compact gauge group is required to be quantised. We temporarily ignore
this requirement as it makes no difference to the present
discussion. It will be more important when non-Abelian gauge fields
and interactions are introduced.

To complete this discussion it is useful to write down the spectrum of
the most general free Lagrangian in (2+1)d involving a single gauge field,
consisting of an arbitrary combination of all three terms: a Maxwell
term, a Chern-Simons term and an explicit mass term. Thus, consider:
\be
{\cal L}= \shalf w\, dA\wedge {}^*dA - 
\shalf x A\wedge dA + \shalf y A\wedge {}^*A
\label{genonefield}
\ee
Here $w,x,y$ are constants. While $w$ must be positive, $x,y$ can
have either sign.

From the preceding discussion we know that for $y=0$ the theory has a
single state of mass $x/w$, and for $w=0$ there is a
single state of mass $-y/x$. For $x=0$ there is a
pair of parity-conjugate states of mass
$\pm\sqrt{y/w}$ (which are tachyonic if $y<0$). 
Finally, a parity transformation $x\to
-x$ should interchange the masses and spins of the two states,
i.e. $m_1\leftrightarrow -m_2$. 
These considerations suffice to determine the spectrum of the
generic theory (which can more directly be obtained by examining
the propagator). We have two states of masses $(m_1,m_2)$ and spins
$(\sign(m_1),\sign(m_2))$ given by:
\be
m_1 = \frac{x+\sqrt{x^2+4wy}}{2w},\quad
m_2 = \frac{x-\sqrt{x^2+4wy}}{2w}
\label{masswxy}
\ee
As $x\to 0$ one gets the expected parity-symmetric answer. For
$w\to 0$ one of the masses goes to infinity and the other to the
desired finite value. Finally for $y\to 0$ we find a massive
state of the correct mass, and apparently an extra massless state.
However in this limit gauge invariance is recovered and has the
effect of decoupling the would-be extra massless state. As a result
there is no massless state in the spectrum of the theory unless {\em
  both} the Chern-Simons {\em and} 
the mass term are absent, i.e. $x=y=0$.

\subsection{Multiple fields}

We now consider multiple 1-form fields in (2+1)d with Chern-Simons and
mass terms. We do not include Maxwell terms to start with. The reason,
as indicated in the introduction, is that we would like to consider
theories with an underlying conformal invariance broken only by a
Higgs expectation value. Such a theory can have Chern-Simons terms and
minimal couplings for the gauge fields, but no Maxwell terms.

Naively one would not expect to find any qualitatively new phenomena
compared to what we have discussed in the previous subsection merely
by introducing additional fields. After all, given a free field theory
we usually diagonalise both the kinetic and mass terms before
introducing interactions, and if this were possible in the presence of
Chern-Simons terms then we would obtain a set of decoupled theories,
each one of the form of \eref{genonefield} with $w=0$. There can be
no massless propagating degrees of freedom in such a theory.

However, for (2+1)d theories with Chern-Simons and mass terms for
multiple spin-1 fields, there is a potential conflict between
diagonalisability of these two terms. As we will see shortly, this can
lead to a qualitatively new feature: the possibility of a massless
propagating state. This is what we call the novel Higgs mechanism
(NHM). There are other interesting features that also arise for the
first time when there are multiple fields.

To understand this issue consider a collection of fields $A^{I},
I=1,2,\cdots n$ described by the most general abelian
Chern-Simons-mass Lagrangian (this can be generalised to include
explicit Maxwell terms):
\be {\cal L}= \shalf k_{IJ}A^{(I)}\wedge dA^{(J)} 
+\shalf m_{IJ} A^{(I)}\wedge
\stardual A^{(J)}
\label{gencs}
\ee
Both $k_{IJ}$ and $m_{IJ}$ are constant real symmetric
matrices. $k_{IJ}$ is taken to be non-degenerate, while $m_{IJ}$ is
allowed to have zero eigenvalues. The transformation:
\be
\delta A^{(I)}=d\Lambda^{(I)}
\ee
is a gauge invariance for every set $\{\Lambda^{(I)}\}$ satisfying
$m_{IJ} \Lambda^{(J)}=0$, i.e. for every null eigenvector of $m$.
Let us now try to bring this action into standard form as a sum of free
actions for $n$ decoupled fields.  

For comparison, we first recall how this is done for a generic free
scalar field theory (in any dimension) with Lagrangian:
\be
-\shalf g_{IJ}\del_\mu \phi^I \del^\mu \phi^J- \shalf
(m^2)_{IJ}\phi^I\phi^J
\ee
where $\phi^i,I=1,2,\cdots,n$ are real scalar fields. Here $g_{IJ}$
and $(m^2)_{IJ}$ are constant real symmetric matrices and $g_{IJ}$ is
positive-definite (otherwise the theory has ghosts). To bring the 
Lagrangian into its
standard form, one first performs an orthogonal
transformation on $\phi^I$ to diagonalise $g_{IJ}$, which then takes
the form ${\rm diag}(g_1,g_2,\cdots,g_n)$ with $g_I>0$ for all $I$.
Next one re-scales the fields:
\be
\phi^I\to\frac{\phi^I}{\sqrt{g_I}}
\ee
so that the kinetic form has the identity metric $\delta_{IJ}$.
Finally one performs another orthogonal transformation on $\phi^I$
that diagonalises $m^2$ while preserving the kinetic term, ending up
with: 
\be -\shalf \del_\mu \phi^I \del^\mu \phi^I- \shalf
m_I^2\phi^I\phi^I 
\ee 
Some of the $m_I$ can be equal to zero and we
can also allow some to be imaginary (i.e. $m_I^2<0$) to allow for
tachyons that are eventually stabilised by a potential. Thus the
theory has been reduced to a collection of decoupled
fields, some massive and others massless.

When we try to apply the analogous procedure to \eref{gencs}, we
find a rather different result. Upon diagonalising $k_{IJ}$, it turns
into ${\rm diag}(k_1,k_2,\cdots,k_n)$ but the eigenvalues $k_i$ are
not required to be positive. The theory with negative eigenvalues, or
both signs of eigenvalues, is perfectly consistent and -- as is now
well-known -- field theories relevant to M-branes
\cite{Bagger:2007jr,Aharony:2008ug} have levels of both signs, which
even permits parity to be conserved\cite{VanRaamsdonk:2008ft}. To be
completely general, we therefore assume there are $p$ negative and $q$
positive eigenvalues with $p+q=n$. Since the $A^{(I)}$ are real, the best
we can do after diagonalising $k_{IJ}$ is to re-scale:
\be
A^{(I)}\to\frac{A^{(I)}}{\sqrt{|k_I|}}
\ee
upon which the action \eref{gencs} reduces to:
\be
{\cal L}= \shalf \eta_{IJ} A^{(I)}\wedge dA^{(J)}
+\shalf m_{IJ} A^{(I)}\wedge
\stardual A^{(J)}
\ee
where $\eta_{IJ}$ is a diagonal matrix with $p$ elements equal to $-1$
and the remaining $q=n-p$ elements equal to $+1$: the
Lorentzian metric preserved by $O(p,q)$.
Hence the linear transformations $A^I\to \Lambda_{IJ}A^J$ 
which preserve the kinetic term are given by matrices
$\Lambda_{IJ}$ satisfying: 
\be
\La^T\eta \La=\eta 
\ee
namely the $O(p,q)$ Lorentz transformations. The mass matrix can
therefore be transformed only as:
\be
m \to \La^T m \La, \La\in O(p,q)
\label{mtrans}
\ee 
In general, a Lorentz transformation is not sufficient to
diagonalise $m$.

We would now like to know under what conditions there exists a Lorentz
transformation that diagonalises $m_{IJ}$ in the basis where $k_{IJ}$
is diagonal. Whenever this is possible, the theory will reduce to a
collection of decoupled free fields with definite masses, and there
will be no new phenomena such as the novel Higgs mechanism. The
transformation law of the matrix $m_{IJ}$ in \eref{mtrans} is that of
a second-rank symmetric tensor under $O(p,q)$ Lorentz transformations.
Therefore this is analogous to the question of whether the
stress-energy tensor $T_{\mu\nu}$ of a field theory can be
diagonalised by Lorentz transformations in a $p+q$-dimensional space
of signature $(p,q)$\footnote{I thank Nemani Suryanarayana for this
  observation and for initial collaboration on the analysis below.}.

We start by considering a general $O(p,q)$ matrix
${\Lambda^I}_J$ which by definition satisfies: 
\be
{\Lambda^I}_K {\Lambda^J}_L\, \eta_{IJ} =
\eta_{KL}, \qquad\eta_{IJ} = {\rm diag} \{ \underbrace{ -1,
  -1, \cdots , -1}_{p}, \underbrace{1, 1, \cdots, 1}_{q}\} 
\label{lametadef}
\ee
where the indices $I,J$ take the $p+q$ values
$-p+1,-p+2,\cdots,-1,0,1,\cdots,q,$ so that non-positive values label
timelike directions. Let us now use $i,j,\cdots$ to label the
space-like directions $(1,2,\cdots q)$, and $m,n,\cdots$ to label the 
time-like directions $(-p+1,-p+2,\cdots, -1,0)$. 
Define a set of $(p+q)$-component vectors 
$V^I_{(m)} = {\Lambda^I}_m$ and another set 
$W^I_{(i)} = {\Lambda^I}_i$. Clearly
these vectors have the following orthonormality properties:
\begin{equation}
\vV_{(m)} \cdot \vV_{(n)} = - \delta_{mn}, ~~ \vW_{(i)} \cdot 
\vW_{(j)} =  \delta_{ij}, ~~ \vV_{(m)} \cdot \vW_{(i)} = 0
\end{equation}
where the inner product is defined using the metric $\eta_{IJ}$.
A collection of vectors $\vV_{(m)}$ and $\vW_{(i)}$
satisfying the above orthogonality relations defines an
element of $O(p,q)$.

If the mass matrix $m_{IJ}$ is block-diagonal, i.e, $m_{im}=0$, then
we can always bring it to a diagonal form using matrices in
$O(p)\times O(q)\subset O(p,q)$. Therefore in order for $m_{IJ}$ to
be diagonalisable, it is sufficient to check whether it can be 
brought into a block-diagonal form. 
There exists an element of $O(p,q)$ which achieves this if and
only if we can find a collection of vectors $\vV_{(m)}$ and $\vW_{(i)}$
such that
\begin{equation}
\label{bdcondition}
V^I_{(m)} W^J_{(i)} \, m_{IJ} = 0 ~~ \forall ~~ m, i \, .
\end{equation}
These then are the diagonalisability conditions. If they are
satisfied the theory breaks up into decoupled free fields, but if not
then it can exhibit more interesting behaviour.

\subsection{Necessary and sufficient conditions: two-field case}
\label{twofield}

Let us first look at a simple example for which $p=q=1$.  We are
working in a basis where the kinetic term has already been
diagonalised and scaled, so $k_{IJ}=(-1,1)$. The most general vectors
$\vV$ and $\vW$ satisfying the conditions above are:
\be
\vV=(\sinh \eta,\cosh\eta), \quad \vW=\pm(\cosh\eta,\sinh\eta)
\ee
Parametrising:
\be
m_{IJ} =\begin{pmatrix} ~a&~b~\\ ~b&~c~\end{pmatrix}
\label{massparam}
\ee
we easily find that \eref{bdcondition} reduces to:
\be
\frac{a+c}{2b}=\coth 2\eta
\ee
from  which the condition for diagonalisability follows:
\be
2\, |b| < |a+c| \, .
\label{diacond}
\ee

The eigenvectors and eigenvalues of $m_{IJ}$ are not
invariant under Lorentz transformations. Hence it is convenient to
reformulate the above condition in terms of the (non-symmetric) 
matrix 
$(\eta\, m)^I_{~J}=\eta^{IK} m_{KJ}$:
\be
(\eta\,m)^I_{~J}=\begin{pmatrix} -a&-b~\\ ~b&~c~\end{pmatrix}
\ee
Since it has one upper and one lower index, this matrix can be thought
of as a linear transformation and one can ask for its eigenvectors and
eigenvalues. These have been classified (in 3+1 dimensions) in works
on general relativity, for example Ref.\cite{Stephani:2003tm}. 
In 1+1 dimensions there are precisely three possibilities:\\[2mm]
\begin{tabular}{l p{6cm} p{8cm}}
&\underline{Eigenvalues} & \underline{Eigenvectors}\\
(i)&Two distinct, real & Two distinct, real (one spacelike, one timelike)\\
(ii)&Two coincident & One\\
(iii)&Complex-conjugate pair & Complex-conjugate pair
\end{tabular}

\vspace*{3mm}

Case (i) allows us to make an $SO(1,1)$ matrix:
\be
\Lambda=\begin{pmatrix} \vv_t& \vv_s\end{pmatrix}
\ee
where $\vv_t,\vv_s$ are the orthonormalised eigenvectors, the first one
timelike and the second spacelike. Clearly $\Lambda$ diagonalises 
$\eta\, m$ by
a similarity transformation:
\be
\Lambda^{-1}\,\eta\, m\,\Lambda = \eta\, m_{\diag}
\ee
where we have labelled the diagonal matrix as $\eta\, m_\diag$. 
Noting that $\Lambda^{-1}=\eta\Lambda^T\eta$, we see that:
\be
\Lambda^T m\Lambda=m_\diag
\ee
as desired. Clearly $\vv_t,\vv_s$ are just the same as $\vV,\vW$.
The other cases do not permit diagonalisation of
$\eta\, m$ and thereby of $m$. 

It is easy to check that the three cases above correspond to three
sets of values for the discriminant of the eigenvalue equation for
$\eta m$, namely $\Delta=(a+c)^2-4b^2$:
\bea
\hbox{(i)}\quad \Delta>0 ~~\to~~ 2|b|&<&|a+c|\nn\\
\hbox{(ii)}\quad \Delta=0 ~~\to~~ 2|b|&=&|a+c|\nn\\
\hbox{(iii)}\quad \Delta <0~~\to~~ 2|b|&>&|a+c|
\label{checkexp}
\eea
Since case (i) admits diagonalisation of the theory, it is trivial. We
therefore study the system for cases (ii) and (iii).  Labelling the
two vector fields as $\Aone,\Atwo$, the Lagrangian is:
\be
\begin{split}
{\cal L}_1 &= -\shalf \Aone\wedge d\Aone + \shalf \Atwo\wedge d\Atwo 
+ \shalf a \Aone\wedge {}^*\Aone\\ & \qquad\quad + b \Aone\wedge {}^*\Atwo 
+ \shalf c \Atwo\wedge {}^*\Atwo
\end{split}
\ee
Note that this Lagrangian is parity conserving, where the parity 
operation is taken as a reflection of space together with  
the interchange $\Aone\leftrightarrow\Atwo$. 

We start with case (ii), namely $|a+c|=2|b|$. As an example 
choose $a=c=\frac{m}{2}, b=-\frac{m}{2}$, so the mass matrix is:
\be
m_{IJ}=\frac{m}{2}
\begin{pmatrix} 
~1 &-1\\ -1 & ~1
\end{pmatrix}
\label{spmassmat}
\ee
Then:
\be
{\cal L}_1 = -\shalf \Aone\wedge d\Aone + \shalf \Atwo\wedge d\Atwo 
+ \sfrac14 m(\Aone-\Atwo)\wedge {}^*(\Aone-\Atwo)
\ee
Since the Chern-Simons and mass terms are not simultaneously
diagonalisable, it is not immediately apparent how to deduce the
spectrum of this theory. However, a field redefinition in terms of even
and odd parity eigenstates:
\be
\begin{split}
C &= \sfrac{1}{\sqrt2}(A^{(2)}+A^{(1)})\nn\\
B &= \sfrac{1}{\sqrt2}(A^{(2)}-A^{(1)})
\end{split}
\ee
casts the Lagrangian into the more useful form:
\be {\cal L}_1 =
B\wedge dC + \shalf m B\wedge \stardual B
\label{novtriv}
\ee 
Now one of the fields, namely $B$, is algebraic. The 
equations of motion are:
\be
\stardual dC=mB,\quad dB=0
\ee
and the first equation can be used to eliminate $B$. Inserting this
back, the Lagrangian reduces to:
\be 
{\cal L}_2=\frac{1}{2m}\, dC\wedge \stardual dC 
\ee 
We see that while ${\cal L}_1$ has the form of a generalised two-field
self-dual theory, ${\cal L}_2$ is instead a {\em massless} Maxwell
Lagrangian. In theories where the mass term of \eref{novtriv} arises
from a Higgs mechanism, what happens physically is that the single
degree of freedom of a real Higgs scalar gets traded for the single
degree of freedom of a massless vector.

Instead of integrating out $B$, an equivalent way to understand
\eref{novtriv} is to define the new 1-form field $B'$:
\be
B'= B- \frac{1}{m}\,\stardual dC
\ee
in terms of which  the action becomes:
\be
{\cal L}'_2= \frac{1}{2m}dC\wedge \stardual dC
+ \shalf m\, B\wedge \stardual B
\ee
and we  have a free Maxwell field plus a decoupled auxiliary field.

The above example is precisely the free-field reduction of the one in
which the NHM was originally discovered\cite{Mukhi:2008ux}. Here we
have derived it from a different point of view: by constructing the
simplest Chern-Simons-mass theory where the Chern-Simons and mass
terms cannot be simultaneously diagonalised.

To study the more general version of case (ii), as well as case (iii),
we continue to work in the basis of definite-parity fields where the
coefficient of the Chern-Simons term is:
\be
k_{IJ}=\begin{pmatrix}~0&~1~\\ ~1&0~\end{pmatrix}
\ee
In this basis the general mass matrix \eref{massparam} becomes:
\be
m_{IJ}= \sfrac12
\begin{pmatrix}
~a+c-2b & -a+c~\\
-a+c & ~a+c+2b~
\end{pmatrix}
\ee
The Lagrangian is then:
\be
{\cal L}=  B\wedge dC+\sfrac14(a+c-2b) B\wedge\stardual B
+\shalf(c-a) B\wedge \stardual C+\sfrac14(a+c+2b)C\wedge \stardual C
\ee
If $a+c\ne 2b$ then the equations of motion can be solved
for $B$. Inserting this solution back into the action, we find:
\be
{\cal L}= \frac{1}{a+c-2b}\Big(dC\wedge \stardual dC
+(a-c) C\wedge dC + (ac-b^2) C\wedge
\stardual C\Big)
\label{newaction}
\ee
It is convenient to change the normalisation of $C$ at this point so
that the coefficient $\frac{1}{a+c-2b}$ of the Lagrangian 
becomes $\shalf$. Then comparing
with \eref{genonefield}, we have:
\be
w= 1,\quad x=c-a,\quad
y = ac-b^2
\ee
It follows from \eref{masswxy} that the spectrum of this
theory generically contains a pair of states of masses:
\be
\shalf\left(c-a\pm\sqrt{(a+c)^2-4b^2}\right)
\ee
As already noted, the spin of the state 
is given by the sign of the mass. For $a=c$ the theory is
parity-conserving since the Chern-Simons term drops out, and there are
two degenerate massive states of spin $\pm 1$ as required by parity.

The theory above has a massless propagating gauge field if and only if
$x=y=0$. This means we must have $a=c=\pm b$. The positive sign is
ruled out because we eliminated the field $B$ on the assumption that
$a+c\ne 2b$ (however there is no loss of generality, since if $a+c=2b$
then we would have eliminated the field $C$ instead and then the
negative sign would have been ruled out). In conclusion, we have shown
that for two fields, the purely Chern-Simons-mass action of
\eref{gencs} with a Lorentzian-signature Chern-Simons term has a
propagating massless mode if and only if $a=c=-b$, in other words
precisely the mass matrix \eref{spmassmat} that we used as an example
of the NHM. This establishes that there is a unique case with two
fields. While we had already seen that the discriminant
$\Delta=(a+c)^2-4b^2$ should be non-positive to ensure
non-diagonalisability of the Lagrangian, we now have a sufficient
condition for NHM that requires both $\Delta=0$ and $a=c$.

There are other interesting cases. Whenever $ac=b^2$, the system has
only a single massive excitation. This can be seen directly from the
Lagrangian where $B$ has been integrated out, namely \eref{newaction},
in which the explicit mass term drops out leaving a topologically
massive theory whose gauge invariance decouples the second excitation.
Another special case is when $|a+c|=\pm 2|b|$ but $a\ne c$. In this
case one finds two massive states of the {\em same} mass and spin.
This may be thought of as ``maximal parity violation'' and represents
another of the interesting situations arising from
non-diagonalisability of the action. While the same spectrum can also
be obtained by just taking a pair of decoupled Chern-Simons-mass
actions, the origin of parity violation is different: in the latter
case it arises from the Chern-Simons terms while in the former case it
comes from the mass terms. This difference will be relevant
after introducing interactions.

Finally, case (iii) in the notation of \eref{checkexp} corresponds to
a discriminant $\Delta<0$. In this case the spectrum consists of a
conjugate pair of {\em complex} masses of spins $\pm 1$. We reserve
judgement on whether such a theory is necessarily inconsistent, since
interactions might conceivably render it consistent. Note that the
fields cannot be redefined to make the masses purely real or imaginary
(tachyonic) because those states would not then be spin eigenstates.
(In the parity conserving case $c=a$, the masses do become purely
imaginary.) Therefore the complex-mass case is one of the
interesting features arising in the spectrum of two-field models.
We also see that the novel Higgs mechanism occurs on
the boundary between a pair of topologically massive Chern-Simons
theories and a Maxwell theory with a complex mass.

\section{General number of fields}

\subsection{Non-diagonalisability: a necessary condition}

There are several different ways to describe the solution to the
conditions for the novel Higgs mechanism in the general situation with
three or more gauge fields. One of these, which follows from
considerations that we analysed in the previous subsection, is the
following. Consider the matrix $m^I_{~J}=\eta^{IK}m_{KJ}$. If this
matrix has $p$ distinct real timelike eigenvectors and $q$ distinct
real spacelike eigenvectors then it can be diagonalised by a
similarity transformation involving an $SO(p,q)$ matrix, otherwise
not. If it is not diagonalisable then we may look for interesting
phenomena including generalisations of the NHM.

The nature of possible eigenvectors of such a matrix in a space of
Lorentzian signature is reviewed in \cite{Stephani:2003tm} (for the
case of 3+1 dimensions) using notation due to Segr\'e and Pleba\'nski.
The case where the matrix possesses a maximal set of nondegenerate
eigenvalues and corresponding distinct eigenvectors is referred to as
{\em algebraically general}. In Segr\'e notation, an algebraically
general tensor has the label $(1\,1\,\cdots 1, 1\,1\,\cdots 1)$ with
$p$ entries before the comma representing timelike eigenvectors and
$q$ entries after the comma representing spacelike eigenvalues. All
other cases are said to be {\em algebraically special}. These then are
the cases for which one has a non-diagonalisable mass matrix. The
novel Higgs mechanism and any other interesting phenomena can
therefore arise only for the algebraically special case.

There is another way of stating the general diagonalisability
condition Ref.\cite{greub,waterhouse} that will be more useful for us.
These works contain a theorem on the possibility of simultaneously
diagonalising a pair of quadratic forms $(A,B)$. In Section 12.12 of
Ref.\cite{greub} it is proved that quadratic forms $A$ and $B$ in more
than two variables can be simultaneously diagonalised by a linear
transformation if they have no common zeros along the diagonal in any
basis\footnote{Notice that we have already encountered the exception
  to this theorem in the two-variable case of the previous
  section. The non-diagonalisable case $\Delta<0$ discussed there 
includes a mass matrix with $a=0$ or $c=0$, in 
a basis where the kinetic matrix is $\diag(-1,1)$. Thus the two
matrices do not have common zeroes along the diagonal but nevertheless
cannot be diagonalised together.}.

For us the two quadratic forms are $k_{IJ}$ and $m_{IJ}$. It is
most convenient to choose a maximally off-diagonal basis for the
former. If we have $p$ timelike and $q$ spacelike directions with
$p<q$ (the analysis is similar for $p\ge q$) we can bring $k_{IJ}$ to
the form: \be k_{IJ}=\begin{pmatrix}
  0& \II_p&0\\
  \II_p & 0&0\\
  0 & 0 & \II_{q-p}
\end{pmatrix}
\label{koffd}
\ee 
From now on we always work in this basis. Then
applying the theorem quoted above, $m_{IJ}$ will be
diagonalisable it does {\em not} have any zeroes on
the diagonal in this basis. This then is the condition 
under which the theory splits into a sum of decoupled 
theories of the form of \eref{genonefield}.

One expects that all other cases, namely those where $m_{IJ}$ has at
least one zero on the diagonal in its first $2p\times 2p$ block, must
represent something more exotic than a collection of decoupled fields.
However that does not mean they all involve massless propagating
fields. We have only found a necessary condition for this, and in the
next section we will look for sufficient conditions.

\subsection{Sufficient conditions: three field case}
\label{threefield}

Let us now consider the case with three fields and a kinetic matrix:
 \be \eta_{IJ}=\begin{pmatrix}
  ~0& ~1&~0\\
  ~1 & ~0&~0\\
  ~0 & ~0 & ~1
\end{pmatrix}
\ee 
The most general Chern-Simons-mass theory with this content has the
Lagrangian: 
\be
\begin{split}
{\cal L} =\,& B\wedge dC + \shalf D\wedge dD + \shalf \alpha B\wedge\stardual B
+ \shalf \beta C\wedge\stardual C + 
\shalf \gamma D\wedge\stardual D\\
&+ \mu B\wedge \stardual C + \nu C\wedge \stardual D + 
\rho D\wedge \stardual B
\end{split}
\label{threefieldlag}
\ee
The equation of motion of $B$ is:
\be
dC + \alpha\, \stardual B + \mu \stardual\, C + \rho \stardual\, D =0
\ee
Solving for $B$ and inserting back in the Lagrangian, we find:
\be
\begin{split}
{\cal L}=\,& \frac{1}{2\alpha} dC\wedge\stardual dC -
\frac{\mu}{\alpha}C\wedge dC +
\shalf\left(\beta-\frac{\mu^2}{\alpha}\right) C\wedge\stardual C
-\frac{\rho}{\alpha}D\wedge dC\\
&+ \left(\nu -\frac{\mu\rho}{\alpha}\right)C\wedge \stardual D
+ \shalf D\wedge dD + \shalf\left(\gamma-\frac{\rho^2}{\alpha}\right)
D\wedge\stardual D
\end{split}
\ee
Thus we have a Maxwell coefficient matrix $Y_{IJ}$, a Chern-Simons
matrix $k_{IJ}$ and a mass matrix $m_{IJ}$ given by:
\be
Y_{IJ} = \begin{pmatrix} 
~\frac{1}{\alpha} &~0\\
~0 &~0 \end{pmatrix},\quad
k_{IJ}= \begin{pmatrix}
-\frac{2\mu}{\alpha}&~-\frac{\rho}{\alpha}\\
-\frac{\rho}{\alpha} & ~~~1
\end{pmatrix},\quad
m_{IJ}= \begin{pmatrix}
~\beta-\frac{\mu^2}{\alpha}& ~~\nu-\frac{\mu\rho}{\alpha}\\
~\nu-\frac{\mu\rho}{\alpha} & ~~\gamma-\frac{\rho^2}{\alpha}
\end{pmatrix}
\label{cdlag}
\ee
Evidently the NHM, leading to a single massless propagating
excitation, arises only if $\beta=\gamma=\mu=\nu=\rho$ so there is no
new example with three fields relative to the case already analysed
for two fields. However the spectrum is in general far more
complicated. The two-field case was previously discussed in
Sec.\ref{twofield}, but with only Chern-Simons and mass terms. Now
that a Maxwell term is also present, it is impossible to diagonalise
all three terms simultaneously in general. Since the Maxwell term goes
away in the far infrared, the analysis for an interacting theory with
the above quadratic terms will follow the analysis of
Sec.\ref{twofield}. For this we need to inspect \eref{cdlag} above
which tells us that the Chern-Simons term has indefinite signature if:
\be
\det k_{IJ} = -\frac{2\mu}{\alpha}- \frac{\rho^2}{\alpha^2}\,<\,0
\ee
Therefore this is the case for which, at least in the infrared, the
Lagrangian cannot be diagonalised but may be analysed following the
procedure outlined above.

\subsection{General case}

The sufficient conditions for NHM in the multi-field case, at least
for $p=q$ (equal number of positive and negative eigenvalues for the
Chern-Simons coefficient) can be found by repeating the procedure for
the two-field case.  We work in the basis where the Chern-Simons
coefficient $k_{IJ}$ is given by \eref{koffd} and divide the $A^{(I)},
I=1,2,\cdots,2p$ into two sets:
\bea
A^i &=& B^i,~i=1,2,\cdots,p\nn\\
A^{p+i} &=& C^i,~i=1,2,\cdots,p\nn 
\eea 
Then the free Chern-Simons-mass Lagrangian takes the form: 
\be 
{\cal L}= B^{i}\wedge dC^{i}
+ \shalf\alpha_{ij} B^i\wedge\stardual B^j +\beta_{ij} B^i\wedge
\stardual C^j +\shalf\gamma_{ij} C^i\wedge \stardual C^j 
\ee
and the corresponding equations of motion are:
\bea
dC^i + \alpha_{ij}\, \stardual B^j + \beta_{ij}\, \stardual C^j&=&0\nn\\
dB^i + \beta_{ij}\, \stardual B^j + \gamma_{ij}\, \stardual C^j&=&0\nn
\eea
Now suppose the matrix $\alpha_{ij}$ is invertible. In that case we
can solve the first equation for $B^i$ and insert this back into the
original Lagrangian to get:
\be
{\cal L}= \shalf\alpha^{-1}_{ij} dC^i\wedge \stardual dC^j
-(\alpha^{-1}\beta)_{ij} C^i\wedge dC^j +
\shalf\left(\gamma-\beta\alpha^{-1}\beta\right)_{ij} C^i\wedge \stardual C^j
\label{suffcond}
\ee
The Chern-Simons term vanishes for every zero eigenvector of $\beta$.
Moreover if such an eigenvector is a simultaneous zero eigenvector of
$\gamma$ then the mass term also vanishes. We conclude that there is
one massless propagating vector field for every simultaneous zero
eigenvector of the matrices $\beta_{ij}$ and $\gamma_{ij}$, under the
condition that $\alpha_{ij}$ is invertible. As in the two-field case,
the roles of $\alpha_{ij}$ and $\gamma_{ij}$ can be interchanged. If
$\alpha_{ij}$ is not invertible, as in a case we will encounter in the
following subsection, then we need that it is nonzero on the common
zero eigenvector of $\beta,\gamma$.

\subsection{Higgs fields}

So far we have been working with generic mass matrices, not specifying
precisely how they arise via a Higgs mechanism. Here we want to ask
which kinds of charged fields give rise to the different types of mass
matrices discussed above. Let us work with an even number $2p$ of
gauge fields. Since we are still working in the Abelian theory, each
Higgs field is a scalar $\phi_\vq$ carrying a set of charges under
these gauge fields. To analyse the physics one needs to fix the basis
for the gauge fields. It is easiest to choose this as in the previous
subsection, i.e. divide them into two sets $B_i,C_i,~i=1,2.\cdots,p$.

Next we choose the charges of a particular Higgs field to be
$\vq=(q_1,q_2,\cdots q_{p}; r_1,r_2\cdots, r_p)$ where the $q_i$ are
charges under $B_i$ and the $r_i$ are charges under $C_i$.  The
kinetic term for such a scalar is:
\be
\Big| \Big(\del_\mu -i\sum_{i=1}^p q_i B_\mu^i
-i\sum_{i=1}^p r_i C_\mu^i
\Big)\phi_\vq\Big|^2
\ee
If $\phi_\vq$ acquires a (complex) vev $\langle
\phi_\vq\rangle=v_\vq$, one gets a mass matrix:
\be
m_{IJ}=2|v_\vq|^2\begin{pmatrix}
q_iq_j & q_i r_j\\
r_iq_j& r_ir_j
\end{pmatrix}
\ee 
We can identify the matrices $\alpha_{ij},\beta_{ij}, \gamma_{ij}$
of the previous subsection with $q_iq_j, q_ir_j, r_ir_j$ respectively
(upto an overall proportionality constant). Now any vector $v_j$
orthogonal to $r_j$, i.e. satisfying $r_jv_j=0$, will be a common zero
eigenvector of the matrices $\beta, \gamma$ and, according to the
analysis of the previous subsection, can give rise to a massless
propagating field. However $q_iq_j$ is also singular,
with only one nonzero eigenvector $q_j$ itself. Therefore to end up
with a massless propagating field requires $q_j v_j\ne 0$. 

As a by-now familiar example, take $p=1$, $N_f=1$, and let the single
scalar carry charges $(q;r)$. The mass matrix is then: 
\be m_{IJ} = 2|v_\vq|^2
\begin{pmatrix}
  q^2 & \,qr\\
  \,qr & r^2
\end{pmatrix}
\ee 
With $q$ or $r$ vanishing, this gives us $a=c=|b|$ in the previous
notation and one has NHM. Thus we have rederived the well-known 
result\cite{Mukhi:2008ux,VanRaamsdonk:2008ft}
that with two gauge fields $A^1,A^2$ (in a diagonal basis for the
Chern-Simons term) and one scalar, the latter must be equally or
oppositely charged under both (``bi-fundamental'', in the non-Abelian
case) to give rise to the novel Higgs mechanism. Moreover, this is
seen to be essentially unique. In the generic case with both $q,r\ne
0$, one easily sees that $|a+c|>2|b|$ so the action is diagonalisable
and there are no novel features.

With a number of independent Higgs fields
$\phi_{\vq^{\,A}}, A=1,2,\cdots,N_f$ we get the mass matrix:
\be
m_{IJ} = 2\sum_{i=1}^{N_f}|v_{\vq^{\,A}}|^2 \begin{pmatrix}
q^A_i q^A_j& q^A_i r^A_j\\
r^A_i q^A_j& r^A_i r^A_j
\end{pmatrix}
\ee 
In this case one can achieve the maximal number of massless
propagating fields, namely $p$, by simply taking $N_f=p$ scalars each
with $q_i\ne 0$ for one $i$ and $r_i=0$ for all $i$. This amounts to a
straightforward set of copies of the two-field bi-fundamental case,
but then one should not expect to find anything more than that at the
Abelian level. Things can become more complicated when there are
non-Abelian interactions. 

\section{Non-abelian case}

The non-Abelian Chern-Simons system with multiple fields presents some
new features that influence the NHM. One well-known feature is that the
coefficient of the Chern-Simons term is quantised for compact gauge
groups and we cannot ignore it or scale it to unity as we did in the
preceding sections.

\subsection{Difference Chern-Simons}

Let us start with the simplest example, a $\cG\times \cG$ theory with
level $k$ for both the gauge groups. We write the difference
Chern-Simons action as:
\be
L_{CS}=\frac{k}{4\pi}\,\tr\left( A\wedge d A 
+ \sfrac23  A\wedge  A\wedge  A -
{\tilde A}\wedge d{\tilde A} - 
\sfrac23 {\tilde A}\wedge {\tilde A}\wedge {\tilde A}\right)
\label{diffcs}
\ee
where $A=A^aT^a$ and $\tr T^a T^b=-\half \delta^{ab}$. For compact $\cG$, 
$k$ is required to be an integer in order to have gauge
invariance under large gauge transformations\cite{Deser:1981wh}.

Comparing with our discussion above, this corresponds to the basis in
which the coefficient matrix $\eta_{IJ}=\diag(-1,1)$. In this basis
the non-Abelian gauge invariance is the normal one:
\be
\delta A = d\Lambda + [A,\Lambda],\qquad
\delta \tA = d\tLambda + [\tA,\tLambda]
\ee

The second basis, in which $\eta_{IJ}$ is purely off-diagonal, is
obtained as before by taking the linear combinations 
\be
 B = \sfrac{1}{2}(A- {\tilde A}),\quad  C = \sfrac{1}{2}
(A + {\tilde A}),\quad
\FC= d C +  C\wedge  C
\ee
(the fields $B$ and $C$ thereby correspond to those in
the Abelian case, but normalised slightly differently to
avoid ugly factors of $\sqrt2$ in the Lagrangian). 
In these variables, the Lagrangian becomes:
\be
L_{CS}=\frac{k}{\pi}\,\tr\Big( B\wedge \FC + 
\sfrac{1}{3}  B\wedge  B\wedge  B\Big)
\label{diffbcvar}
\ee
and the gauge transformations take the somewhat non-standard form:
\bea
\delta B &=& d\Lambda_B + [C,\Lambda_B] + [B,\Lambda_C]\nn\\ 
\delta C &=& d\Lambda_C + [C,\Lambda_C] + [B,\Lambda_B]
\label{bctransf}
\eea

The equations of motion in the $A,\tA$ basis are just the equations
for two independent flat connections:
\be
F=\tF=0
\ee
However in the $B,C$ basis they look like:
\bea
\FC+B\wedge B &=& 0\nn\\
\DC B \equiv dB +[C,B]&=& 0 
\label{eqnom}
\eea 
Now we would like to give a mass via the Higgs mechanism. The
structure of the mass matrix arises via the choice of representation
of the group $\cG\times \cG$ under which the Higgs field
transforms. The canonical choice is the {\em bi-fundamental}, for
example the $(N,{\bar N}$ of $SU(N)$:
\be
\delta \Phi = -\Lambda \Phi + \Phi\tLambda
\ee
In  this case the covariant derivative is:
\be
D_\mu\Phi = \del_\mu \Phi +A_\mu\Phi - \Phi \tA_\mu
\ee
For convenience we will normalise the scalar kinetic term as:
\be
\frac{k}{4\pi}\tr (D_\mu\Phi^\dagger D^\mu\Phi)
\ee
where this trace is, formally, unrelated to that in the gauge field
action -- here it just sums over two pairs of repeated indices in the
fundamental representation, one pair being associated to each gauge group 
of ${\cal G}\times {\cal G}$. This kinetic term gives rise to the
interaction:
\be 
\frac{k}{4\pi}\tr\Big|A_\mu\Phi - \Phi \tA_\mu\Big|^2
\ee
With a Higgs vev proportional to the identity: $\langle \Phi\rangle =
v\II$, the mass term is equal to 
\be
\frac{k}{4\pi}v^2\tr(A_\mu-\tA_\mu)^2
\ee 
where now the trace is over the Lie algebra of ${\cal G}$ after
identifying the two factors in ${\cal G}\times {\cal G}$.
We see that the mass matrix has the form:
\be
m_{IJ}\sim 
\begin{pmatrix}
~~1 & -1\\
-1 & ~~1
\end{pmatrix}
\label{massneg}
\ee
which we have encountered many times over as the basic example of
NHM. 

This time, however, one has to keep track of the cubic terms in $B$.
With the above vev, one has the Lagrangian:
\be
{\cal L}=\frac{k}{\pi} \tr\left(B\wedge F^{(C)} + \sfrac{1}{3}B\wedge B\wedge
  B - v^2 B\wedge \stardual B
\right)
\label{cublag}
\ee 
From this follows the equation of motion for $B$:
\be
\FC + B\wedge B - 2 v^2\,\stardual B=0
\ee
The quadratic term in $B$ now makes it impossible, unlike in the
Abelian case, to simply solve for $B$ and eliminate it. The best we
can do is solve recursively, to get:
\be
\begin{split}
B&= -\frac{1}{2v^2}\stardual\FC -\frac{1}{2v^2}\stardual (B\wedge B)\\
&= -\frac{1}{2v^2}\stardual\FC -\frac{1}{8v^6}\stardual 
(\stardual\FC\wedge \stardual\FC)
  + \cdots
\end{split}
\label{recurs}
\ee
The terms in $\cdots$ above contain all powers of $\FC$, appearing in
combinations like:
\be
\stardual (\stardual \FC\wedge \stardual
(\stardual \FC\wedge\cdots \stardual(\stardual
\FC\wedge\stardual \FC)))
\label{multifc}
\ee
These are made by wedging the 1-form dual to $\FC$ with
itself, then dualising the result back to a 1-form and repeating
indefinitely. This nonlinear combination of field strengths appears to
be unique to (2+1)d. Moreover, the orders in this expansion are
counted by the parameter $\frac{1}{v^2}$.

We may now insert \eref{recurs} back into the Lagrangian of
\eref{cublag} to find:
\be
{\cal L} = \frac{k}{\pi}\left(-\frac{1}{4v^2}\FC\wedge\stardual \FC
-\frac{1}{24v^6}\stardual \FC\wedge\stardual \FC\wedge\stardual \FC
+\cdots
\right)
\ee
We see that taking $v\to\infty$ allows us to ignore the higher-order
terms in $\FC$. However, in the quadratic term, $v^2$ plays the role
of the Yang-Mills coupling constant\cite{Mukhi:2008ux}.  Therefore in
the same limit that decouples the higher-order terms, the Yang-Mills
term becomes very strongly coupled. This can be avoided by
simultaneously scaling $k\to\infty,v\to\infty$ keeping $\frac{k}{v^2}$
fixed\cite{Distler:2008mk}. In this limit the higher-order terms do
drop out, but the Yang-Mills coupling $\frac{v}{\sqrt{k}}$ remains
finite and can be chosen arbitrarily.

One can take the Higgs field in a slightly different representation
that is equivalent to the previous one in the Abelian case but differs in the
non-Abelian case. This is the bi-fundamental $(N,N)$ rather than 
$(N,{\bar N})$\footnote{This representation 
arises in the fermionic sector of
  certain orientifold field theories, as in Ref.\cite{Armoni:2008kr}.}. 
In this case the gauge transformation is:
\be
\delta\Phi = -\Lambda\Phi - \Phi\tLambda
\ee
and the covariant derivative is:
\be
D_\mu\Phi = \del_\mu \Phi +A_\mu\Phi + \Phi \tA_\mu
\ee
This time the vev $\langle \Phi\rangle =
v\II$ gives rise to a mass matrix:
\be
m_{IJ}\sim 
\begin{pmatrix}
~1 & ~1\\
~1 & ~1
\end{pmatrix}
\ee
which at the Abelian level has the same properties as \eref{massneg}
above and gives rise to the NHM, with the difference that the field $C\sim
A+\tA$ is eliminated and the field $B\sim A-\tA$ becomes propagating. 
However in the non-Abelian case the cubic terms
are different. The Lagrangian after Higgsing is:
\be
{\cal L}=\frac{k}{\pi} \tr\left(B\wedge F^{(C)} + \sfrac{1}{3}B\wedge B\wedge
  B - v^2 C\wedge \stardual C
\right)
\ee 
We need to integrate out $C$, so the first term has to be rewritten:
\be
\begin{split}
B\wedge \FC &= B\wedge dC + B\wedge C\wedge C\\
&= C\wedge dB + B\wedge C\wedge C
\end{split}
\ee
where we have performed integration by parts on the kinetic term. The
equation of motion for $C$ is:
\be
dB + 2B\wedge C -2v^2\, \stardual C=0
\ee
which gives:
\be
\begin{split}
C&= -\frac{1}{2v^2}\,\stardual dB -\frac{1}{v^2}\,\stardual (B\wedge
C)\\
&=-\frac{1}{2v^2}\,\stardual dB +\frac{1}{2v^4}\,\stardual (B\wedge
\stardual dB)
+\cdots
\end{split}
\ee
This time the higher-order terms are made up of combinations like:
\be
\stardual (B\wedge 
\stardual (B\wedge\cdots \stardual
(B\wedge\stardual dB)))
\label{multib}
\ee
which should be contrasted with \eref{multifc}. 

It is manifest from the construction of \eref{multifc} and
\eref{multib} that these nonlinear terms are gauge invariant (after
taking a trace) under the transformations in \eref{bctransf}. These
(perhaps novel) gauge-invariant combinations are building blocks that
may have interesting applications in the study of M-theory membrane
actions.

One can consider yet other representations for the Higgs field(s). But
as in the Abelian case, it is clear that fields which transform only
under one or other gauge group give uninteresting decoupled gauge
symmetries.

\subsection{$B\wedge F$ theory}

There is a different non-Abelian generalisation of the Abelian
difference Chern-Simons theory.  Instead of writing a difference of
two non-Abelian Chern-Simons terms, we start directly with
the action:
\be 
2\,\tr B\wedge \FC 
\label{bcaction}
\ee
with gauge invariances: 
\bea
\delta B&=& d\Lambda_B + [ C,\Lambda_B]+[B,\Lambda_C]\nn\\
\delta C&=&d\Lambda_C+[ C,\Lambda_C]
\eea 
At the Abelian level this is still the difference of two
Chern-Simons actions. However the absence of the last term in $\delta
C$ as compared with \eref{bctransf} means the non-Abelian gauge
invariance is different. Introducing a suitable scalar
field one can get a mass term: 
\be v^2\, \tr\, B_\mu B^\mu 
\label{bmass}
\ee
Because of the absence of a $B^3$ term, this time the end-point of the
NHM is the massless Yang-Mills theory {\em without} any
corrections. This means the theory above can be thought of as a
reformulation of (2+1)d Yang-Mills theory, a point which was stressed
in the M-theory context in Ref.\cite{Ezhuthachan:2008ch}.

The theory in \eref{bcaction} together with the mass term in
\eref{bmass} can be obtained by starting with \eref{cublag},
performing the re-scaling: 
\be 
B\to \lambda^{-1} B,\quad k\to \lambda
k,\quad v\to \sqrt\lambda{v} 
\ee
and then taking the limit
$\lambda\to\infty$. In this sense it is just a limiting case of
difference Chern-Simons theory. Note that the final Yang-Mills
coupling $\gYM^2\sim v^2/k$ remains finite in this limit. It is not
clear that $k$ should be quantised in the final theory, in fact it can
be absorbed in a redefinition of $B$.

\section{Potential applications}

In this section I offer some speculative ideas about the possible
occurrence of the NHM in the context of condensed-matter physics as
well as gravity.  Though there will be no definite conclusion in
either case, it is tantalising that known field theories arising in
both contexts quite naturally exhibit a structure of multiple gauge
fields and a difference-Chern-Simons Lagrangian. Therefore it is
plausible that, with more investigation, a physical setting for the
NHM can be found in each case.

\subsection{Condensed-matter systems}

Abelian Chern-Simons gauge fields play a key role in understanding
phenomena like the
fractional quantum Hall effect (see for example
Refs.\cite{Semenoff:1988jr,Zhang:1988wy,Read:1988mp,Wen:1991xx}). 
In particular
the Chern-Simons term generates a change from Bose/Fermi to anyon
statistics. Several gauge fields appear in the problem. To start with
there is of course the electromagnetic field, which being massless has
only a Maxwell term. Subsequently a ``statistical gauge field'' is
introduced, having a Chern-Simons term with a suitable coefficient. In
the treatment of Ref.\cite{Wen:1991xx} there
are also several other gauge fields. A set of $m$ gauge fields is
introduced so that the fermionic operators in the Lagrangian can be
replaced by bosonic operators, one for each of $m$ Landau levels.
Another set comes in upon dualising the Goldstone modes into gauge
fields using scalar-gauge duality in (2+1)d.

Let us briefly review this and see how structures similar to those we
have been discussing in previous sections could arise. 
In units where $\hbar=c=e=1$, $e$ being the electric charge quantum, a
fractional Hall system in some definite Landau level is described by
the Lagrangian (the electrons $\psi$ are nonrelativistic):
\be
{\cal L} = \psi^\dagger\, i(\del_0-iA_0)\psi + \frac{1}{2m}\psi^\dagger
(\del_i-iA_i)^2\psi -\frac14 F_{\mu\nu}F^{\mu\nu}
\ee
where $A_\mu$ is the external electromagnetic field. The filling
fraction is $\nu=\frac{2\pi n}{B}$ where $n$ is the number of
electrons per unit area and $B$ is the magnetic field, $B=dA$.

We now modify this Lagrangian by introducing a new Abelian gauge field
$\alpha_\mu$ and modify the above Lagrangian to:
\be
\begin{split}
{\cal L} &= \psi^\dagger\, i\Big(\del_0-i(A_0+\alpha_0)\Big)\psi 
+ \frac{1}{2m}\psi^\dagger
\Big(\del_i-i(A_i+\alpha_i)\Big)^2\psi\\
&\qquad\quad +\frac12 dA\wedge \stardual dA +
\frac{1}{4\pi p} \alpha\wedge d\alpha
\end{split}
\ee
The equation of motion for $\alpha_\mu$ is:
\be
\stardual d\alpha = -2\pi p j
\ee
where $j_\mu = \Big(\psi^\dagger \psi, ~\psi^\dagger \vec\nabla
\psi\Big)$ is the current. It follows that the electrons effectively
move in a reduced magnetic field $B_{\eff}= B-2\pi pn$. Thereby the
filling fraction changes to:
\be
\nu~\to~ \nu_{\eff} = \frac{2\pi n}{B_{\eff}}=\frac{\nu}{1-p\nu}
\ee
If the original filling fraction is of the form $\nu=\frac{m}{mp+1}$
with $m$ an integer, then we find $\nu_{\eff}=m$. 
The quantity $p$ also contributes to the effective statistics of the
electrons, which now have $p$ flux lines attached to them
providing a phase of $e^{i\pi p}$. Therefore if we want to retain Fermi
statistics, $p$ should be a (positive or negative) even integer. The
net result is that we now have a system of fermions in the reduced
magnetic field exhibiting the {\em integer} quantum Hall 
effect\cite{PhysRevLett.63.199,Greiter:1990kn}. The
field $\alpha$ is called the ``statistical gauge field'' for reasons
which should be clear.

The next step in Ref.\cite{Wen:1991xx} is to replicate the above
system over $m$ different Landau levels, with a Fermi field $\psi_I$
for each level. Then, introduce a new set of
$m$ gauge fields $a_{I\mu}, ~I=1,2,\cdots,m$ with a suitable
Chern-Simons term that converts each fermion field $\psi_I$ to a boson
$\phi_I$. At this stage the Lagrangian is:
\be
\begin{split}
{\cal L}&= \sum_{I=1}^{m} \Bigg[ 
\phi_I^\dagger\, i\Big(\del_0-i(A_0+\alpha_0+a_{I0})\Big)\phi_I
+\frac{1}{2M}
\phi_I^\dagger\, \Big(\del_i-i(A_i+\alpha_i+a_{Ii})\Big)^2\phi_I\\
&\qquad\qquad -V_I(\phi_I^\dagger\phi_I) + \frac{1}{4\pi}a_I\wedge da_I
\Bigg]+ \half dA\wedge \stardual dA
+ \frac{1}{4\pi p}\alpha\wedge d\alpha 
\end{split}
\ee
where a potential $V_I(\phi_I^\dagger \phi_I)$ has been introduced
to represent the interaction between electrons in the corresponding
Landau level.

For simplicity let us set $m=1$. Ignoring the physical
electromagnetic field for the moment, the gauge-field terms in the
above Lagrangian are:
\be
\begin{split}
{\cal L}&= \phi^\dagger\, i\Big(\del_0-i(\alpha_0+a_0)\Big)\phi
+\frac{1}{2M}\phi^\dagger\, \Big(\del_i-i(\alpha_i+a_i)\Big)^2\phi\\
&\qquad\qquad + \frac{1}{4\pi}a\wedge da
+ \frac{1}{4\pi p}\alpha\wedge d\alpha 
\end{split}
\ee
This looks something like the two-field system we analysed earlier, or
rather a non-relativistic version of it. With a Higgs vev, the mass
matrix will acquire the form:
\be
m_{IJ}\sim \begin{pmatrix}
~1&~1~\\ ~1&~1~\end{pmatrix}
\ee
The physics now depends on the parameter $p$. If $p$ is positive then
the mass term can be diagonalised, one eigenvalue is zero and the
other nonzero, leading to one field remaining non-dynamical and the
other being massive and propagating. If $p$ is negative then the
Chern-Simons kinetic matrix has indefinite signature and the action is
potentially non-diagonalisable. In this case, after scaling fields so
that the kinetic matrix becomes $\diag(-1,1)$ the mass matrix takes
the form:
\be
m_{IJ}\sim \begin{pmatrix}
~1&\sqrt{|p|}\\ \sqrt{|p|}&~|p|\end{pmatrix}
\ee
For any even $p$ this always satisfies the constraint in
\eref{diacond} and therefore the Lagrangian is again diagonalisable
and leads to the same spectrum. This is satisfying because in the FQHE
system the physics does not appear to depend significantly on the sign
of the Chern-Simons terms, which are introduced for the sake of
statistics.

It remains to consider $p=-1$, a case that was not allowed in the FQHE
context. Here we indeed encounter the novel Higgs
mechanism\footnote{In the present context, perhaps it is 
better called the ``novel Anderson-Higgs mechanism''...} and the
spectrum has a single propagating massless mode. However we need to
examine whether the analysis is valid in the presence of the external
gauge field $A$, which has a Maxwell kinetic term but mixes in the
mass matrix with both the Chern-Simons fields. The resulting theory
actually resembles the 3-field case analysed in Section
\ref{threefield}, with the relabelling $(B_\mu,C_\mu)\to 
(a_\mu\pm\alpha_\mu)$ and $D_\mu\to A_\mu$. With this, the 3-field 
mass matrix in \eref{threefieldlag} becomes:
\be
\begin{pmatrix}
~\alpha &\mu &\rho~\\
~\mu &\beta & \nu~\\
~\rho &\nu &\gamma~
\end{pmatrix}
~~\to~~
\sim v^2 \begin{pmatrix}
~2 &~0 &~1~\\
~0 &~0 & ~0~\\
~1 &~0 &~1~
\end{pmatrix}
\ee
so $\mu=\nu=\beta=0$. After eliminating $B_\mu$, we find for the analogue
of \eref{cdlag} the following matrices (here we have taken account of
the fact that the field $A_\mu$ has a Maxwell rather than Chern-Simons
kinetic term):
\be
Y_{IJ} = \begin{pmatrix} 
~\frac{1}{2v^2} &~0\\
~0 &~1 \end{pmatrix},\quad
k_{IJ}= \begin{pmatrix}
0&-\frac12\\
-\frac12 & ~0
\end{pmatrix},\quad
m_{IJ}= v^2\begin{pmatrix}
~0& ~0\\
~0 & ~\half
\end{pmatrix}
\label{cdlagtwo}
\ee

The above matrices are written in the basis of fields $(C_\mu,A_\mu)$.
In the absence of the electromagnetic field $A_\mu$ we would retain
only the top-left corner of these matrices and would find, as
discussed above, that $C_\mu=a_\mu-\alpha_\mu$ is a massless
propagating field. However we now see that the presence of $A_\mu$
complicates the theory considerably. In fact, elimination of $B_\mu$
has induced a Chern-Simons term for the electromagnetic field $A_\mu$
and the Chern-Simons kinetic matrix above is of Minkowski signature,
so it cannot be diagonalised simultaneously with the mass matrix.
Matters are complicated by the presence of Maxwell terms for both
fields, but if it can be justified to ignore them at long distances
then we would have a fresh novel Higgs mechanism leaving one
combination of the electromagnetic field $A_\mu$ and
$C_\mu=a_\mu-\alpha_\mu$ as a massless propagating field.

The purpose of this sub-section has not been to present and solve for
a specific condensed-matter system exhibiting NHM (in particular,
recall that the discussion above was carried out for $p=-1$ while $p$
is supposed to be even in the given systems). Instead it has been 
pointed out that the kind of theories discussed in the context of the
quantum Hall effect, and more generally theories in which vortices and
unusual statistics play an important role, do have multiple
Chern-Simons terms and these can be mutually non-diagonalisable with
the mass terms, making them plausible settings for the NHM. The study
of what precise physical effect this induces, and in which
specific many-body system, is left for future work.

\subsection{(2+1)d gravity}

It is known\cite{Achucarro:1987vz,Witten:1988hc} that in 2+1 dimensions,
gravity can be written as a difference-Chern-Simons theory. One takes
the variables to be the dreibein $e_\mu^{~a}$ and $\omega_\mu^{~ab}$
as is usual in the first-order formalism, and combines them into a
pair of 1-forms:
\be
A_\mu^a = \half \epsilon^a_{~bc}\omega_\mu^{~bc}+
\frac{1}{l}e_\mu^{~a},\qquad
\tA_\mu^a = \half \epsilon^a_{~bc}\omega_\mu^{~bc}-
\frac{1}{l}e_\mu^{~a}
\label{eomegaforms}
\ee
where $l$ is a constant with dimensions of length.
Taking $T^a,a=1,2,3$ to be the generators of SL(2,R), normalised as 
$\tr(T^aT^b)=\half\eta^{ab}$, and defining $A_\mu = A_\mu^a T^a$, the
Lagrangian:
\be
{\cal L} = \frac{l}{16\pi G_N}\tr\Big(
A\wedge dA+\sfrac{2}{3}A\wedge A\wedge A
-\tA\wedge d\tA-\sfrac{2}{3}\tA\wedge \tA\wedge \tA\Big)
\ee
can be shown to be equivalent to ordinary Einstein gravity in (2+1)d
with a negative cosmological constant $\Lambda=-\frac{3}{l^2}$.

Except for the choice of a specific non-compact gauge group, SL(2,R),
this is identical to the Lagrangian \eref{diffcs} that we discussed in
Section 4.1. It is also natural to write this action in the
form of \eref{diffbcvar} because the fields $B_\mu, C_\mu$ defined
there are, in the present case:
\be
B_\mu^a = \frac{1}{l}e_\mu^{~a},\qquad C_\mu^a = \half 
\epsilon^a_{~bc}\omega_\mu^{~bc}
\ee
This gives a nice physical interpretation in the context of gravity,
to the fields $B_\mu,C_\mu$ of Section 4.1. 

We also see that in terms of these fields, the terms $B\wedge dC$ and
$B\wedge C\wedge C$ are of order $\frac{1}{l}$ while the term $B\wedge
B\wedge B$ is of order $\frac{1}{l^3}$. Thus the limit $l\to\infty$ is
easily taken and leads to the pure $B\wedge \FC$ Lagrangian
\eref{bcaction} of Sec. 4.2, with now $B_\mu^a = e_\mu^{~a}$ since the
$l$-dependence has cancelled against the coefficient of the Lagrangian. This
Lagrangian therefore corresponds to (2+1)d gravity in the absence of a
cosmological constant.

It is worth remarking that there is a beautiful generalisation of the
above structure to the case of higher-spin 
fields\cite{Henneaux:2010xg,Campoleoni:2010zq}. Here the 1-forms in
\eref{eomegaforms} are generalised to include the corresponding
quantities describing one or more higher-spin fields. With this
generalisation all the above formulae carry over identically.

It remains to ask whether there is a novel Higgs mechanism in this
system. This would require a mass term: $\sim-\tr (B_\mu B^\mu)=-\half
\eta^{\mu\nu}\eta_{ab}e_\mu^{~a}e_\nu^{~b}$ which is of course not
generally covariant. In fact, in the Chern-Simons formulation general
covariance is replaced by SL(2,R) $\times$ SL(2,R) gauge invariance,
and a mass term will not be gauge-invariant, so this is not
necessarily a surprise.  However, an additional problem is that the
metric $\eta_{ab}$ of SL(2,R) has Lorentzian signature so the above
``mass term'' for $e_\mu^{~a}$ has wrong signs for some components (in
our metric, the right sign would be negative for spacelike values of
$\mu$, but here we get positive signs when $\mu$ is spacelike and $a$
is timelike). Finally, if such a term were nevertheless generated and
formally used to integrate out $e_\mu^{~a}$, the result would be
SL(2,R) Yang-Mills theory which by itself continues to be plagued
with sign issues.

It is still (barely) conceivable that a mass term for $e_\mu^{~a}$ is
generated along just one (spacelike) direction of SL(2,R). This
would Higgs a U(1) $\times$ U(1) part of the Chern-Simons action and
lead to a Maxwell kinetic term. We leave this possibility for a future
investigation.

\section*{Acknowledgements}

I would like to thank Guillaume Bossard, Kedar Damle, Bobby
Ezhuthachan, Rajesh Gopakumar, Kimyeong Lee, Shiraz Minwalla, Nitin
Nitsure, Costis Papageorgakis, David Tong and particularly Nemani
Suryanarayana for helpful discussions. Generous support for the basic
sciences by the people of India is gratefully acknowledged.

\section*{Notation and conventions}

We work with the (2+1)d metric: 
\be
\eta_{\mu\nu}={\rm diag} (-, +, +)
\ee
The differential form notation used throughout is easily translated
into conventional index notation using the following identites: 
\be
\begin{split}
A\wedge \stardual A &= -A_\mu A^\mu\\
A\wedge dA&=\epsilon^{\mu\nu\lambda}A_\mu \del_\nu A_\lambda\\
dA\wedge \stardual dA &= -\shalf F_{\mu\nu}F^{\mu\nu}
\end{split}
\ee
Finally, in  the non-Abelian context with compact groups, we have:
\be
A=A^a T^a, F=F^a T^a
\ee
where $T^a$ are anti-Hermitian, and:
\be
\tr\, T^a T^b = -\half \delta^{ab}
\ee

\bibliographystyle{JHEP}
\bibliography{CS-Higgs}

\providecommand{\href}[2]{#2}\begingroup\raggedright\begin{thebibliography}{10}

\bibitem{Jackiw:1980kv}
R.~Jackiw and S.~Templeton, {\it {How superrenormalizable interactions cure
  their infrared divergences}},  {\em Phys.Rev.} {\bf D23} (1981) 2291.

\bibitem{Schonfeld:1980kb}
J.~F. Schonfeld, {\it {A mass term for three-dimensional gauge fields}},  {\em
  Nucl.Phys.} {\bf B185} (1981) 157.

\bibitem{Deser:1981wh}
S.~Deser, R.~Jackiw, and S.~Templeton, {\it {Topologically massive gauge
  theories}},  {\em Ann. Phys.} {\bf 140} (1982) 372--411.

\bibitem{Mukhi:2008ux}
S.~Mukhi and C.~Papageorgakis, {\it {M2 to D2}},  {\em JHEP} {\bf 05} (2008)
  085, [\href{http://xxx.lanl.gov/abs/0803.3218}{{\tt arXiv:0803.3218}}].

\bibitem{Semenoff:1988jr}
G.~W. Semenoff, {\it {Canonical quantum field theory with exotic statistics}},
  {\em Phys.Rev.Lett.} {\bf 61} (1988) 517.

\bibitem{Zhang:1988wy}
S.~Zhang, T.~Hansson, and S.~Kivelson, {\it {An effective field theory model
  for the fractional quantum Hall effect}},  {\em Phys.Rev.Lett.} {\bf 62}
  (1988) 82--85.

\bibitem{Read:1988mp}
N.~Read, {\it {Order parameter and Ginzburg-Landau theory for the fractional
  quantum Hall effect}},  {\em Phys.Rev.Lett.} {\bf 62} (1989) 86--89.

\bibitem{Wen:1991xx}
X.~G. Wen and A.~Zee, {\it Topological structures, universality classes, and
  statistics screening in the anyon superfluid},  {\em Phys. Rev. B} {\bf 44}
  (1991) 274--284.

\bibitem{Bandres:2008vf}
M.~A. Bandres, A.~E. Lipstein, and J.~H. Schwarz, {\it {N = 8 superconformal
  Chern--Simons theories}},  {\em JHEP} {\bf 05} (2008) 025,
  [\href{http://xxx.lanl.gov/abs/0803.3242}{{\tt arXiv:0803.3242}}].

\bibitem{VanRaamsdonk:2008ft}
M.~Van~Raamsdonk, {\it {Comments on the Bagger-Lambert theory and multiple M2-
  branes}},  {\em JHEP} {\bf 05} (2008) 105,
  [\href{http://xxx.lanl.gov/abs/0803.3803}{{\tt arXiv:0803.3803}}].

\bibitem{Deser:1984kw}
S.~Deser and R.~Jackiw, {\it {`Selfduality' of topologically massive gauge
  theories}},  {\em Phys. Lett.} {\bf B139} (1984) 371.

\bibitem{Townsend:1983xs}
P.~K. Townsend, K.~Pilch, and P.~van Nieuwenhuizen, {\it {Selfduality in odd
  dimensions}},  {\em Phys. Lett.} {\bf 136B} (1984) 38.

\bibitem{Deser:1989gf}
S.~Deser and Z.~Yang, {\it {A remark on the Higgs effect in presence of
  Chern-Simons terms}},  {\em Mod. Phys. Lett.} {\bf A4} (1989) 2123.

\bibitem{Bagger:2007jr}
J.~Bagger and N.~Lambert, {\it {Gauge symmetry and supersymmetry of multiple
  M2-branes}},  {\em Phys. Rev.} {\bf D77} (2008) 065008,
  [\href{http://xxx.lanl.gov/abs/0711.0955}{{\tt arXiv:0711.0955}}].

\bibitem{Aharony:2008ug}
O.~Aharony, O.~Bergman, D.~L. Jafferis, and J.~Maldacena, {\it {N=6
  superconformal Chern-Simons-matter theories, M2-branes and their gravity
  duals}},  {\em JHEP} {\bf 10} (2008) 091,
  [\href{http://xxx.lanl.gov/abs/0806.1218}{{\tt arXiv:0806.1218}}].

\bibitem{Stephani:2003tm}
H.~Stephani {\em et.~al.}, {\em {Exact solutions of Einstein's field
  equations}}.
\newblock Cambridge University Press, UK, 2003.

\bibitem{greub}
W.~Greub, {\em {Linear Algebra}}.
\newblock Springer-Verlag, 1963.

\bibitem{waterhouse}
W.~C. Waterhouse, {\it {Pairs of quadratic forms}},  {\em Inv. Math.} {\bf 37}
  (1976) 157.

\bibitem{Distler:2008mk}
J.~Distler, S.~Mukhi, C.~Papageorgakis, and M.~Van~Raamsdonk, {\it {M2-branes
  on M-folds}},  {\em JHEP} {\bf 05} (2008) 038,
  [\href{http://xxx.lanl.gov/abs/0804.1256}{{\tt arXiv:0804.1256}}].

\bibitem{Armoni:2008kr}
A.~Armoni and A.~Naqvi, {\it {A non-supersymmetric large-N 3D CFT and its
  gravity dual}},  {\em JHEP} {\bf 0809} (2008) 119,
  [\href{http://xxx.lanl.gov/abs/0806.4068}{{\tt arXiv:0806.4068}}].

\bibitem{Ezhuthachan:2008ch}
B.~Ezhuthachan, S.~Mukhi, and C.~Papageorgakis, {\it {D2 to D2}},  {\em JHEP}
  {\bf 07} (2008) 041, [\href{http://xxx.lanl.gov/abs/0806.1639}{{\tt
  arXiv:0806.1639}}].

\bibitem{PhysRevLett.63.199}
J.~K. Jain, {\it Composite-fermion approach for the fractional quantum hall
  effect},  {\em Phys. Rev. Lett.} {\bf 63} (1989) 199--202.

\bibitem{Greiter:1990kn}
M.~Greiter and F.~Wilczek, {\it {Heuristic principle for quantized Hall
  states}},  {\em Mod.Phys.Lett.} {\bf B4} (1990) 1063--1070.

\bibitem{Achucarro:1987vz}
A.~Achucarro and P.~Townsend, {\it {A Chern-Simons action for three-dimensional
  anti-de Sitter supergravity theories}},  {\em Phys.Lett.} {\bf B180} (1986)
  89.

\bibitem{Witten:1988hc}
E.~Witten, {\it {(2+1)-dimensional gravity as an exactly soluble system}},
  {\em Nucl.Phys.} {\bf B311} (1988) 46.

\bibitem{Henneaux:2010xg}
M.~Henneaux and S.-J. Rey, {\it {Nonlinear $W_{infinity}$ as asymptotic
  symmetry of three-dimensional higher spin anti-de Sitter gravity}},  {\em
  JHEP} {\bf 1012} (2010) 007, [\href{http://xxx.lanl.gov/abs/1008.4579}{{\tt
  arXiv:1008.4579}}].

\bibitem{Campoleoni:2010zq}
A.~Campoleoni, S.~Fredenhagen, S.~Pfenninger, and S.~Theisen, {\it {Asymptotic
  symmetries of three-dimensional gravity coupled to higher-spin fields}},
  {\em JHEP} {\bf 1011} (2010) 007,
  [\href{http://xxx.lanl.gov/abs/1008.4744}{{\tt arXiv:1008.4744}}].

\end{thebibliography}\endgroup

\end{document}